# Excitation of multiple surface plasmon-polaritons by a metal layer inserted in an equichiral sculptured thin film

SH. Hosseininezhad, F. Babaei [*]

Department of Physics, University of Qom, Qom, Iran

*)Email: fbabaei@qom.ac.ir

## Abstract

Excitation of multiple surface plasmon-polaritons(SPPs) by an equichiral sculptured thin film with a metal layer defect was studied theoretically in the Sarid configuration, using the transfer matrix method. Multiple SPP modes were distinguished from waveguide modes in optical absorption for p- polarized plane wave. The degree of localization of multiple SPP waves was investigated by calculation of the time averaged Poynting vector. The results showed that the long-range and short-range SPP waves can simultaneously be excited at both interfaces of metal core in this proposed structure which may be used in a broad range of sensing applications.

## I. Introduction

Multiple surface plasmon-polaritons(SPPs) have attracted more attention due to potential applications in sensing, imaging and optical communications [1-4]. Sculptured thin films are two and three dimensional artificial structures that can support the excitation of multiple SPPs [5,6]. These films may be produced by using oblique angle deposition and substrate rotation around axes perpendicular and parallel to the substrate as chiral and nematic morphology, respectively [7,8]. In traditional chiral sculptured thin films, the substrate continuously rotates about its surface normal [9]. However, if the substrate holder includes the discrete abrupt angular rotations of $2\pi/Q$, a Reusch pile (a polygonal chiral thin film) can be formed [10,11] ,where the parameter Q>2 is an integer. A Reusch pile with Q=4 is called equichiral sculptured thin film (ECSTF) and other is named ambichiral sculptured thin film [12].

It is well known that only one SPP wave can be excited at single interface of thin metal film and homogenous dielectric material [13]. The propagation length of an SPP wave is short due to metal damping loss. Then, the high SPP wave damping owing to ohmic losses in metal limits its propagation length to the order of a few micrometers [14]. One of the practical geometries for enhancement of the propagation length and more localization of the SPP wave at metal-dielectric interface is the Sarid configuration [15]. In this configuration, a metal thin film is surrounded by two dielectric materials, while the set is placed on the high index prism as a trilayer composite. The bottom, top of dielectric layers and metal thin film are respectively named substrate, superstrate(cover) and guide medium [16]. Thus, there are two substrate/metal and metal/superstrate interfaces that two SPP waves can propagate on them. For thick metal film, only one SPP wave can be excited at metal / substrate interface, while a moderate thickness of the metal film includes two SPP modes, one at substrate/metal interface and other at metal/superstrate interface [12]. In a thin metal film, two SPP modes can be hybridized as symmetric and anti-symmetric wave. The symmetric and anti-symmetric SPP waves respectively are known as long- range SPP(LRSPP) and short -range SPP(SRSPP) waves. The propagation length of the LRSPP wave is at least a factor of 2 to 3 higher than that of the single interface SPP wave, resulting in propagation over a longer distance [17].

Before the presentation of the sculptured thin films by Lakhtakia and Messier at meeting held at Penn State [18,19], the possibility of excitation of multiple SPPs was introduced using the multiple parallel metal-isotropic dielectric interfaces [20]. When a vertically periodic nonhomogeneous dielectric material be deposited on metal thin film, the multiple SPP waves can be excited at single metal-dielectric interface [21]. A sculptured thin film is periodically nonhomogeneous normal to the interface, then can excite multiple SPP modes at single interface in the vicinity of a homogeneous metal [22]. The phenomenon of multiple SPP modes has been named surface multiplasmonics [13]. They can be used in sensors for simultaneous detection of more than one chemical species [2,4].

In this work, an ECSTF with a central metal thin layer defect is considered theoretically in the Sarid configuration. Optical absorption for p-polarized plane wave as a function of incident angle of light at different vapor deposition and offset angles for different thicknesses of metallic layer obtained by the use of the transfer matrix method. Multiple SPP peaks were extracted from the absorption plots and also the angular position of them has been tested with the change in structural period(pitch) of the bottom and top of ECSTF. In order to analysis of the hybridization of multiple SPP modes, Cartesian components of the time-averaged Poynting vector of them along the thickness of the structure have been reported. In the simulation model the following parameters are used. An $\exp(i\omega t)$ time dependence is implicit, with $\omega$ denoting the angular frequency. The free-space wave number, the free-space wavelength, and the intrinsic impedance of free space are denoted by $k_0 = \omega\sqrt{\varepsilon_0\mu_0}$, $\lambda_0 = 2\pi/k_0$ and $\eta_0 = \sqrt{\mu_0/\varepsilon_0}$, respectively, with $\mu_0$ and $\varepsilon_0$ being the permeability and permittivity of free space, respectively. Vectors are underlined once, dyadics are underlined twice, the asterisk denotes the complex conjugate and Cartesian unit vectors are identified as $\underline{u}_{x,y,z}$. The theoretical formulation is given in Section II and followed by the results and discussion in Section III.

## II. Theoretical formulation

Suppose that the region $-d_{met}/2 \leq z \leq d_{met}/2$ is occupied by an isotropic homogenous metal with relative permittivity $\varepsilon_{met}$ (Fig.1). The regions $-(d_b + (d_{met}/2)) \leq z \leq -(d_{met}/2)$ and $(d_{met}/2) \leq z \leq (d_t + (d_{met}/2))$ are respectively considered as the bottom and top of ECSTF, while the regions $z \leq -(d_b + (d_{met}/2))$ and $z \geq (d_t + (d_{met}/2))$ are respectively occupied by a high index prism with relative permittivity $\varepsilon_1 = n_1^2$ and an air medium with refractive index $n_2 = 1$. Let us assume that this set in the name of the Sarid configuration can be exposed by a linear polarized plane wave from the bottom of the prism at an angle $\theta_{inc}$ to the z- axis. The phasors of incident, reflected and transmitted electric fields are given:

$$\underline{E}_{inc}(\underline{r}) = [\underline{S} a_S + \underline{P}_{inc} a_P] e^{i[k\,x + k_0\,n_1\,(z+d_b+\frac{d_{met}}{2})\cos\theta_{inc}]}, \quad z \leq -(d_b + \frac{d_{met}}{2})$$

$$\underline{E}_{ref}(\underline{r}) = [\underline{S} r_S + \underline{P}_{ref} r_P] e^{i[k\,x - k_0\,n_1\,(z+d_b+\frac{d_{met}}{2})\cos\theta_{inc}]}, \quad z \leq -(d_b + \frac{d_{met}}{2}) \quad (1)$$

$$\underline{E}_{tr}(\underline{r}) = [\underline{S} t_S + \underline{P}_{tr} t_P] e^{i[k\,x + k_0\,n_2(z-d_t-\frac{d_{met}}{2})\cos\theta_{tr}]}, \quad z \geq (d_t + \frac{d_{met}}{2})$$

The phasor of the magnetic field in any region is given as $\underline{H}(\underline{r}) = (i\omega\mu_0)^{-1} \underline{\nabla} \times \underline{E}(\underline{r})$, where $(a_S, a_P), (r_S, r_P)$ and $(t_S, t_P)$ are the amplitudes of incident plane wave, and reflected and transmitted waves with s- and p-linear polarizations and also $k = k_0 n_1 \sin\theta_{inc}$. The unit vectors for linear polarization normal and parallel to the incident plane, $\underline{S}$ and $\underline{P}_{inc,ref,tr}$, respectively are defined as:

$$\underline{S} = \underline{u}_y$$
$$\underline{P}_{inc} = -\underline{u}_x \cos\theta_{inc} + \underline{u}_z \sin\theta_{inc}$$
$$\underline{P}_{ref} = \underline{u}_x \cos\theta_{inc} + \underline{u}_z \sin\theta_{inc} \quad (2)$$
$$\underline{P}_{tr} = -\underline{u}_x \cos\theta_{tr} + \underline{u}_z \sin\theta_{tr}$$

, where $\sin\theta_{tr} = (n_1/n_2)\sin\theta_{inc}$, $\cos\theta_{tr} = +\sqrt{1 - \sin^2\theta_{tr}}$.

The reflectance and transmittance amplitudes can be obtained, using the continuity of the tangential components of electrical and magnetic fields at four interfaces of trilayer composite and solving the algebraic matrix equation:

$$\begin{bmatrix} t_S \\ t_P \\ 0 \\ 0 \end{bmatrix} = [\underline{\underline{K}}(\theta_{tr})]^{-1} [\underline{\underline{M}}_t][\underline{\underline{M}}_m][\underline{\underline{M}}_b][\underline{\underline{K}}(\theta_{inc})] \begin{bmatrix} a_S \\ a_P \\ r_S \\ r_P \end{bmatrix} \qquad (3),$$

The transfer matrix of any region and other terms of this equation are given in detail in Appendix A.

Finally, using the obtained amplitudes in Eq. 3, then we can calculate the reflectance and transmittance coefficients as $r_{ij} = \dfrac{r_j}{a_i}$ and $t_{ij} = \dfrac{t_j}{a_i}$; $i, j = s, p$. The optical absorption for s- and p- linear polarizations is calculated as $A_i = 1 - \sum_{j=s,p} R_{ji} + T_{ji}, i = s, p$, where $R_{ij} = |r_{ij}|^2$ (reflection) and $T_{ij} = (n_2 \operatorname{Re}(\cos\theta_{tr})/(n_1 \cos\theta_{inc}))|t_{ij}|^2$ (transmission); $i, j = s, p$ so that Re( ) is the real part of the quantity given in the parenthesis.

### III. Numerical results and discussion

For purpose of the simulation, the structure of the trilayer composite is considered as a bulk aluminum layer (as a defect) that it is inserted in a right-handed TiO$_2$ ECSTF. In all calculations, the free space wavelength $\lambda_0 = 633\, nm$ and the dielectric permittivity of the metal $\varepsilon_{met} = -56 + 21i$ were fixed. The pitch of ECSTF was fixed as $2\Omega = 300\ nm$ and the dielectric permittivity of the prism was $\varepsilon_1 = 6.656$ (ZnSe) [23]. In fact the ECSTF is anisotropic, the SPP wave can be excited by both s- and p- polarized plane wave but the number of SPP waves here is too much we only considered a p-polarized plane wave for all results presented. A Mathematica program was written for solve Eq.3 to obtain optical absorption. In our calculations, the plasmon peaks that its optical absorption intensity is less than 0.4 are not considered (although under 0.4 value there is the several SPP modes but we only considered the SPP peaks with high optical absorption). Because, it is difficult in detection of nanoparticles adsorbed on surface in sensed medium at low intensity of optical absorption. The relative permittivity scalars for TiO$_2$ ECSTF at $\lambda_0 = 633\,nm$ are assumed as [24]:

$$\varepsilon_a(\chi_v) = [1.0443 + 2.7394(\frac{2\chi_v}{\pi}) - 1.3697(\frac{2\chi_v}{\pi})^2]^2$$
$$\varepsilon_b(\chi_v) = [1.6765 + 1.5649(\frac{2\chi_v}{\pi}) - 0.7825(\frac{2\chi_v}{\pi})^2]^2 \quad (4),$$
$$\varepsilon_c(\chi_v) = [1.3586 + 2.1109(\frac{2\chi_v}{\pi}) - 1.0554(\frac{2\chi_v}{\pi})^2]^2$$

where $\chi_v$ is the vapor deposition angle, the relation between the tilt angle ($\chi$) of nanocolumns of ECSTF and $\chi_v$ is as $\tan\chi = 2.8818\tan\chi_v$.

Multiple SPP peaks extract from optical absorption versus γ (with increment 30°) for different vapor deposition angles $\chi_v \in \{5°,15°,25,°35,°45°,60°\}$ with a d$_{met}$ =10 nm are shown in Fig.2. We found two SPP peaks for $\chi_v = 5°$ that their average angular position($\tilde{\theta}$) are located at $23.07°(\tilde{\theta}_{1SPP})$ and $28.63°(\tilde{\theta}_{2SPP})$. It is observed that at $\chi_v = 15°$ there is a continuity for $\theta_{1SPP}$ and $\theta_{2SPP}$ with $\tilde{\theta}_{1SPP} = 26.60°$ and $\tilde{\theta}_{2SPP} = 35.91°$. Four SPP modes can be clearly distinguished for $\chi_v = 25,35,45°$ so that their average angular position are respectively at $\tilde{\theta}_{1,2,3,4SPP} = 22.90, 32.20, 40.09, 48.17°$ for $\chi_v = 25°$, $\tilde{\theta}_{1,2,3,4SPP} = 22.88, 38.05, 44.66, 54.51°$ for $\chi_v = 35°$ and $\tilde{\theta}_{1,2,3,4SPP} = 22.85, 43.15, 49.13, 59.86°$ for $\chi_v = 45°$. Also, there exist three SPP for $\chi_v = 60°$ modes with $\tilde{\theta}_{1,2,3SPP} = 23.12, 48.91, 66.43°$. One easily can obtain the wavenumbers of SPP waves from the average angular position of them as $K_{SPP} = K_0 n_1 \sin\theta_{SPP} + i K_0 n_1 \sin\Gamma_{SPP}$ [12,16] ,where $\Gamma$ is the full width at half maximum (FWHM). It is expected that the several SPP peaks at interface of a metal and a nonhomogeneous dielectric can be excited. In here, ECSTF at the vapor deposition angles 0° and 90° must be closed to air and TiO$_2$ homogenous isotropic mediums, respectively. The principal relative permittivity scalars $\varepsilon_{a,b,c}$ for homogenous isotropic materials are identical( $\varepsilon_a = \varepsilon_b = \varepsilon_c$) but this matter cannot be obtained from Eq.4( $\varepsilon_a = 1.09, \varepsilon_b = 2.81, \varepsilon_c = 1.85$ at $\chi_v = 0°$ and $\varepsilon_a = 5.83, \varepsilon_b = 6.05, \varepsilon_c = 5.83$ at $\chi_v = 90°$). The validity of this equation is in the range of $\chi_v \in [7°, 90°]$[12]. In fact, the relationship between $\chi_v$ and $\chi$ is more complicated and modified for silicon-oxide and magnesium-fluoride columnar thin films[25,26]. Therefore, the number of SPP modes must be reduced with close to $\chi_v = 0, 90°$ due to homogeneity and isotropy.

Fig.3 shows the angular position all plasmonic modes as a function of γ for different vapor deposition angles with a d$_{met}$ =20 nm. It is seen that there are two continuity SPP modes for $\chi_v = 5°$, the first vary almost slightly and smoothly versus γ with $\tilde{\theta}_{1SPP} = 22.91°$ and the second is included a rippling wavelike with $\tilde{\theta}_{2SPP} = 29.13°$. Three branches of SPP are obtained for $\chi_v = 15°, 60°$ with $\tilde{\theta}_{SPP} = 26.02°, 35.80°, 40.19°$, $\tilde{\theta}_{SPP} = 22.99°, 47.78°, 69.15°$ ,respectively so that they continuously vary in terms of offset angle. Also, at $\chi_v = 35°, 45°$ three SPP modes appear, while the first and third branch vary linearly versus γ with $\tilde{\theta}_{1SPP} = 37.14°, 42.08°$, $\tilde{\theta}_{3SPP} = 55.95°, 61.79°$ and the second mode with $\tilde{\theta}_{2SPP} = 44.52°, 49.09°$ is discontinuous. When $\chi_v = 25°$, three continuous SPP modes occur and one SPP mode discontinuous appears, so that the average angular position of the former is $\tilde{\theta}_{1,2,4SPP} = 22.91°, 31.44°, 49.01°$ and the latter is $\tilde{\theta}_{3SPP} = 39.98°$. The calculations were also performed for a $d_{met}$=30 nm that the obtained results are given in Fig.4. At $\chi_v \in \{5°, 45°\}$ the values of $\theta_{SPP}$ contain two branches, while the first mode discontinuity vary versus γ with $\tilde{\theta}_{1SPP} \in \{23.13°, 42.22°\}$ and the second mode vary almost slightly and smoothly with $\tilde{\theta}_{2SPP} \in \{29.19°, 63.14°\}$. When $\chi_v \in \{15°, 25°, 35°\}$ exist only one continuity branch of SPP with $\tilde{\theta}_{1SPP} \in \{40.68°, 49.74°, 56.98°\}$, respectively. In our work for $\chi_v = 60°$ there exist two continuity of SPP modes and one discontinuity branch of SPP so that the average angular position of the second, third and first are located at $\tilde{\theta}_{2,3SPP} \in \{47.71°, 71.05°\}$ and $\tilde{\theta}_{1SPP} = 22.95°$, respectively. Generally, a comparison between Figs.2,3&4 reveals that for thin metal ($d_{met}$=10 nm) the number of SPP modes is more than $d_{met}$=20,30 nm. Because, as metal thickness is less than skin depth of metal (for bulk aluminum at $\lambda_0$=633 nm is 13.24 nm), some of branches of SPP are propagated at ECSTF/metal interface (external interface) which may be hybridized with those are excited at metal/ECSTF interface (internal interface) as a symmetric or antisymmetric SPP wave [21]. As expected, when the metal thickness exceeds the skin depth, the coupling of SPP modes between two interfaces decreases and for very thick metal SPP waves may be guided only by the metal/ECSTF interface [12].

Absorption $Ap_{Nb\,Nt}$ as a function of incident angle of light for different vapor deposition and offset angles are depicted in Fig.5 so that Nb and Nt are respectively the period of bottom and top of the ECSTF, when $d_{met}$= 10 nm. It can be seen that the multiple modes (absorption peaks) appear in absorption plots, but all of them do not indicate to an excitation of SPP at an interface. The position of SPPs (as a function of incident angle of light) does not depend on the thickness of dielectric medium (in here, thickness of ECSTF); while in the other peaks that are named waveguide modes, their position changes with variations of thickness of dielectric medium [1,5,21, 22]. In simulations, we tested all multiple modes as a function of dielectric medium and only SPP modes are reported which the calculated results are shown in Fig.2.For this purpose, one period(pitch=2Ω) of bottom and top of the ECSTF (Nb=Nt=1) was considered as a reference structure. A time, one period added to bottom of ECSTF and next time, one period added to top of ECSTF. The optical absorption of these three structure was drawn together in a plot and then the angular position of SPP peaks was extracted. In Fig.5, only the obtained results in some of offset angles are presented here.

Fig.6 shows reflection and transmission spectra versus $\theta_{inc}$ for some of different vapor deposition and offset angles, when $d_{met}$ = 20 nm and $N_d$ =$N_t$ =1 except that it is mentioned(Fig.6c). It is obvious that the values of transmission are lower than reflection. Multiple modes are located in troughs of reflection spectra. On the other hand, multiple SPP peaks $\theta_{SPP}$ can be obtained from reflection using depiction of $Rp_{11}$, $Rp_{12}$ & $Rp_{21}$. Again, if in these plots the position of a trough is fixed by changing of ECSTF thickness, it is a SPP mode otherwise is a wave guide mode that only for a sample is shown in Fig.6c. The SPP troughs are exactly the same the SPP peaks in terms of angular position. Also, multiple SPP peaks versus γ extracted from the optical absorption (Ap) at $\chi_v = 90°$ for $d_{met}$ =10,20,30 nm and absorption, reflection and transmission of them as a function of incident angle of light at γ = 90,180,270°,respectively are drawn in Fig.7. We found three SPP branches for $d_{met}$ =10,20 nm and two SPP branches for $d_{met}$=30 nm(Figs.7a-7c). There is no dependence on γ because the ECSTF is isotropic in the x-y plane for $\chi_v = 90°$ ($\varepsilon_a = \varepsilon_c = 5.83, \varepsilon_b = 6.05$ ).Although, for an ECSTF/metal interface there exist only one branch for all values of γ[22] but herein more than one SPP occur that may be caused by the presence of two interfaces(ECSTF/metal and metal/ECSTF) and the hybrid SPP modes at them.Figs.7d-7f simultaneously show absorption, reflection and transmission spectra at γ=90,180,270° for $d_{met}$=10,20,30 nm, respectively. It is seen that the values of SPP peaks in absorption are coincident on the values of SPP troughs in reflection. Therefore, the existence of a sharp peak in optical absorption versus $θ_{inc}$ with linear polarization indicates to excitation a SPP wave which is equivalent with a sharp trough in reflection spectra without a compensatory peak in transmission plot [27-29].

Up to now, the SPP modes from the waveguide modes as it was expressed in Fig.5 have been distinguished but it must be known that these modes in which one of the interfaces are propagated or may be hybridized together. In order to more analysis, Cartesian components of the time-averaged Poynting vector $P(z)=(1/2)\text{Re}[E\times H^*]$ versus the thickness of the trilayer composite at some different vapor deposition and offset angles for $d_{met}$ =10 nm are depicted in Fig.8, when a SPP wave has been excited. It is evident that the strong hybridization between SPP waves can be occurred at ECSTF/metal and metal /ECSTF interfaces in Figs.8b4,8c3&8d3. In fact, these SPP modes are excited at higher incident angles of light which are more localized, include sharp absorption peak and have low full width half maximum(FWHM) that are consistent with Figs.2&5. Also, the relative phase speed($(1/n_1 \sin θ_{SPP})$) of these SPP wave are less than rest modes[30,31]. The strong coupling leads to increase the propagation length of SPP and then they are classified in LRSPP waves. It is seen that the coupling between two interfaces decreases in Figs.8b2,8c2,8d1&8d2 so that it can be considered these SPP waves only propagate on the ECSTF/metal interface. Although a weak hybridization is observed but the intensity of $P_x(z)$ at external of interface is high, then we called these SPP in here as SRSPP waves. In Fig.8b1, the SPP wave only can be excited at internal interface as a SRSPP. A medium coupling between internal and external interfaces exist in Figs.8a1&8c1 which the behavior of $P_x(z)$ curve is wavelike. In Figs.8a2&8b3 there is a semi anisotropic hybrid SPP wave so that it is more localized at external and internal interfaces, respectively.

The similar calculations were repeated for metal thickness 20 nm and 30 nm in Fig.9 and Fig.10, respectively. We found that there exists the strong confinement of SPP modes at both interfaces of metal in Figs.9a3,9b3,9c3&9d3.This more localization of SPP waves is caused that a strong hybridization between them occurs. In Figs.9a1&9d1 a medium coupling between both interfaces of metal happens that its field propagates versus thickness of structure as a semi symmetric SPP wave. The weak coupling between internal and external interfaces occurs in Figs.9b1,9c1&9d2 that is more localized at ECSTF/metal interface. In Figs.9b2&9c2 the SRPP wave is bounded to metal/ECSTF interface. An antisymmetric SPP wave can be excited in Fig.9a2 at metal/ECSTF interface. Almost the same results were found in Fig.10 with the difference that they are more localized at internal interface due to thick metal thickness. Also herein, in Figs.10a2,10b1&10c1 and Figs.10d1,10e2,10f3&10g2 are available the strong hybridization of SPP modes at both interfaces of metal that the former and the later can be propagated as symmetric and antisymmetric waves, respectively. A medium coupling between internal and external interfaces as semi symmetric SPP wave occurs in Figs.10e1,10f2&10g1 which it has short propagation length. The curve of $P_x(z)$ in Fig.10f1shows a wavelike SPP wave in bottom of ECSTF that it is more localized at internal interface. In Fig.10a1, SPP wave can be excited only at metal/ECSTF interface.

## IV.Conclusion

The excitation and hybridization of multiple SPP waves from an ECSTF with a central metal layer defect were investigated theoretically in the Sarid configuration by using the transfer matrix method. The optical absorption was drawn for p-polarized plane wave as a function of incident of light at different vapor deposition and offset angles. The multiple SPP modes were distinguished from the waveguide modes by adding one structural period (pitch) to bottom and top of ECSTF. We found that there exist the multiple SPP peaks so that some of them can be propagated as continuity or discontinuity versus offset angle. The results showed that as metal thickness increases the number of SPP modes decreases. The LRSPP waves occur at higher incident of light for a fixed vapor deposition angle which can be strongly hybridized at both interfaces of metal. The strong hybrid modes include a symmetric SPP wave, although the other SPP modes occur as a semi symmetric or semi antisymmetric SPP wave. The SRSPP and medium range SPP waves appear at low incident of light that the former only can be bounded to internal or external interface of metal and the later relatively moderate can be coupled to both interfaces of metal. Also, it is found that the multiple SPP waves only can be excited at internal interface of metal for a thick metal. Finally, in this trilayer composite there exist simultaneously the LRSPP and SRSPP waves so that each of them has the potential applications in the felid of plasmonics.

**Acknowledgements**

The authors would like to express their deep gratitude to the University of Qom and Iran National Science Foundation (INSF) for supporting this work.

## Appendix A: The transfer matrix method

The transfer matrix of the bottom, top of ECSTFs and metallic thin film can be obtained by solving the source-free Maxwell curl postulates in that regions:

$$\nabla \times \underline{E}(\underline{r}) = i\omega \underline{B}(\underline{r})$$
$$\nabla \times \underline{H}(\underline{r}) = -i\omega \underline{D}(\underline{r}) \quad (A1),$$

where D and B are the displacement electric and the magnetic induction fields, respectively. The constitutive relations for D and B are as:

$$\underline{D}(\underline{r}) = \varepsilon_0 \underline{\underline{\varepsilon}}_{ECSTF} \cdot \underline{E}(\underline{r})$$
$$\underline{B}(\underline{r}) = \mu_0 \underline{H}(\underline{r}) \quad (A2) \quad \text{in } l \text{ th arm of the ECSTF and}$$

$$\underline{D}(\underline{r}) = \varepsilon_0 \varepsilon_{met} \underline{E}(\underline{r})$$
$$\underline{B}(\underline{r}) = \mu_0 \underline{H}(\underline{r}) \quad (A3) \quad \text{in}$$

metal mediums. The $\varepsilon_{met}$ is the metal homogenous dielectric permittivity and the nonhomogeneous dielectric permittivity $\underline{\underline{\varepsilon}}_{ECSTF}$ for $l$ th arm of the ECSTF is defined as[12]:

$$\underline{\underline{\varepsilon}}_{ECSTF} = \underline{\underline{S}}_z^T(\zeta) \cdot \underline{\underline{S}}_y^T(\chi) \cdot \underline{\underline{\varepsilon}}_{ref} \cdot \underline{\underline{S}}_y^T(\chi) \cdot \underline{\underline{S}}_z(\zeta) \quad (A4),$$

where the superscript T indicates to the transpose of a dyadic, $\zeta = h(l-1)(2\pi/Q) - \gamma$, $\chi$ is tilt angle of the nanocolumns of the ECSTF, $h = +1 (or -1)$ is right-handedness (or left-handedness) of the ECSTF and $\gamma$ is the offset angle( the starting point for growth) of the nanocolumns of the ECSTF relative to the x-axis in x-y plane. In our work we fixed $\psi_{inc} = 0°$ and the structure rotates in clockwise direction so that it is considered as $-\gamma$ in the $\zeta$ angle(Fig.1).

The local relative permittivity, rotation and tilt dyadics are respectively as [32]:

$$\underline{\underline{\varepsilon}}_{ref} = \varepsilon_a \underline{u}_z \underline{u}_z + \varepsilon_b \underline{u}_x \underline{u}_x + \varepsilon_c \underline{u}_y \underline{u}_y$$
$$\underline{\underline{S}}_z = (\underline{u}_x \underline{u}_x + \underline{u}_y \underline{u}_y) \cos \zeta + (\underline{u}_y \underline{u}_x - \underline{u}_x \underline{u}_y) \sin \zeta + \underline{u}_z \underline{u}_z \quad (A5),$$
$$\underline{\underline{S}}_y = (\underline{u}_x \underline{u}_x + \underline{u}_z \underline{u}_z) \cos \chi + (\underline{u}_z \underline{u}_x - \underline{u}_x \underline{u}_z) \sin \chi + \underline{u}_y \underline{u}_y$$

where $\varepsilon_{a,b,c}$ are the relative permittivity scalars.

With considering electromagnetic fields in dielectric and metal regions are as follows:

$$\underline{E}(\underline{r}) = \underline{e}(z) e^{ikx}$$
$$\underline{H}(\underline{r}) = \underline{h}(z) e^{ikx} \quad (A6),$$

and using the Eqs.A1-A3, one can obtain four ordinary differential equations and two algebraic equations. The two algebraic equations for $e_z$ and $h_z$ in ECSTF(for $l$ th arm ) and metal mediums are respectively as follows:

$$e_z(z) = \frac{\frac{(\varepsilon_a - \varepsilon_b)}{2}\sin 2\chi\, A - \frac{k}{\varepsilon_0 \omega} h_y(z)}{B}$$

$$h_z(z) = \frac{k}{\mu_0 \omega} e_y(z)$$

(A7),

where $A = (e_x(z)\cos\zeta - e_y(z)\sin\zeta)$ and $B = (\varepsilon_a \cos^2\chi + \varepsilon_b \sin^2\chi)$.

$$e_z(z) = -\frac{k}{\varepsilon_0 \omega \varepsilon_{met}} h_y(z)$$

$$h_z(z) = \frac{k}{\mu_0 \omega} e_y(z)$$

(A8).

The four ordinary differential equations can be sorted as a matrix equation:

$$\frac{d}{dz}[\underline{f}(z)] = i[\underline{\underline{P}}(z)][\underline{f}(z)] \tag{A9},$$

where $[\underline{f}(z)] = [e_x(z)\ e_y(z)\ h_x(z)\ h_y(z)]^T$ is a column vector. The $[\underline{\underline{P}}(z)]$ is a $4\times 4$ matrix and the elements of it for $l$ th arm of the ECSTF are :

$$P_{11} = \frac{k(\varepsilon_a - \varepsilon_b)\cos\zeta \sin 2\chi}{2B},\ P_{12} = -P_{11}\tan\zeta,\ P_{14} = \mu_0\omega - \frac{k^2}{\varepsilon_0 \omega B},\ P_{23} = -\mu_0\omega,$$

$$P_{31} = \varepsilon_0\omega\frac{\sin 2\zeta}{2}(\frac{\varepsilon_a \varepsilon_b}{B} - \varepsilon_c),\ P_{32} = \frac{k^2}{\mu_0 \omega} - \varepsilon_0\omega(\frac{\varepsilon_a \varepsilon_b \sin^2\zeta}{B} + \varepsilon_c \cos^2\zeta),$$

$$p_{34} = -P_{12},\ P_{41} = \varepsilon_0\omega(\frac{\varepsilon_a \varepsilon_b \cos^2\zeta}{B} + \varepsilon_c \sin^2\zeta),\ P_{42} = -P_{31},\ P_{44} = P_{11}$$

(A10), and the other elements are zero. The elements of $[\underline{\underline{P}}(z)]$ in metal region are :

$$P_{14} = \mu_0\omega - \frac{k^2}{\varepsilon_0 \omega \varepsilon_{met}},\ P_{23} = -\mu_0\omega,\ P_{32} = \frac{k^2}{\mu_0 \omega} - \varepsilon_0\omega\varepsilon_{met},\ P_{41} = \varepsilon_0\omega\varepsilon_{met} \tag{A11},$$

and the else elements are zero.

We now consider an ECSTF with thickness of $d = Nt$, where $t$ is the thickness of an arm of ECSTF and N is the number of arms. The transfer matrix of a columnar thin film with thickness of t is $e^{i[\underline{\underline{P}}]t}$. Therefore, the transfer matrix of an ECSTF is [30,31]:

$$[\underline{\underline{M}}] = [\underline{\underline{M}}]_N [\underline{\underline{M}}]_{N-1} \cdots [\underline{\underline{M}}]_3 [\underline{\underline{M}}]_2 [\underline{\underline{M}}]_1 \quad \text{(A12)},$$

where $[\underline{\underline{M}}]_l = e^{i[\underline{\underline{P}}_l]t}, l = 1, 2, \ldots, N$. Then, this method can be used to obtain the transfer matrix of the bottom and top of ECSTFs in Eq.3 ($[\underline{\underline{M}}_b]$ and $[\underline{\underline{M}}_t]$). Also, easily the transfer matrix of metal is just $[\underline{\underline{M}}]_m = e^{i[\underline{\underline{P}}]d_{met}}$.

By rewriting Eqs.1 as $[\underline{f}(z = -(d_b + \frac{d_{met}}{2})_-] = [\underline{\underline{K}}(\theta_{inc})][a_S \ a_P \ r_S \ r_P]^T$ in incident medium (including incident and reflected electric fields) and $[\underline{f}(z = (d_t + \frac{d_{met}}{2})_+] = [\underline{\underline{K}}(\theta_{tr})][t_S \ t_P \ 0 \ 0]^T$ in transmitted medium, the elements of $[\underline{\underline{K}}(\theta_{inc})]$ are:

$$K_{12} = -K_{14} = -\cos\theta_{inc}, \ K_{21} = K_{23} = 1, \ K_{31} = -K_{33} = \frac{n_1}{\eta_0} K_{12}, \ K_{42} = K_{44} = \frac{-n_1}{\eta_0} \quad \text{(A13)}$$

and the elements of $[\underline{\underline{K}}(\theta_{tr})]$ are:

$$K_{12} = -K_{14} = -\cos\theta_{tr}, \ K_{21} = K_{23} = 1, \ K_{31} = -K_{33} = \frac{n_2}{\eta_0} K_{12}, \ K_{42} = K_{44} = \frac{-n_2}{\eta_0} \quad \text{(A14)}.$$

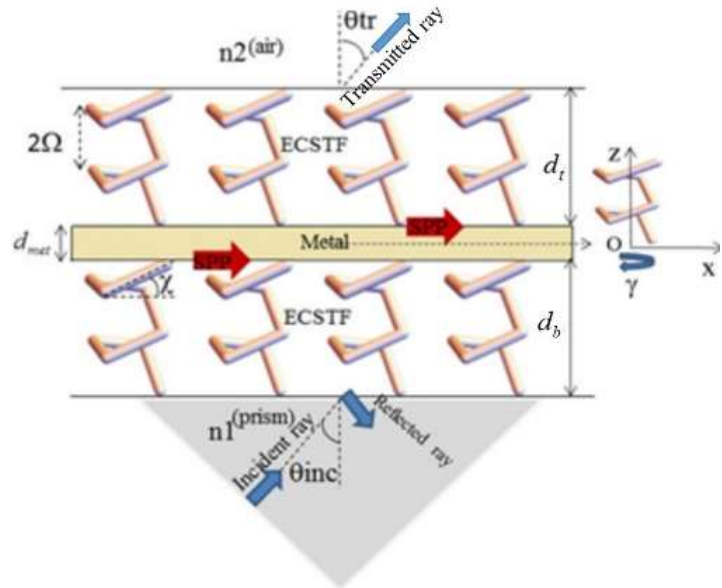

**Fig.1.** Schematic illustration of the geometry of the an ECSTF with a central metallic layer defect in Sarid configuration for the excitation of multiple SPP. The $\Omega$, $\chi$ and $\gamma$ are respectively the half pitch, tilt angle and the offset angle (the starting point for growth) of nanocolumns in ECSTF.

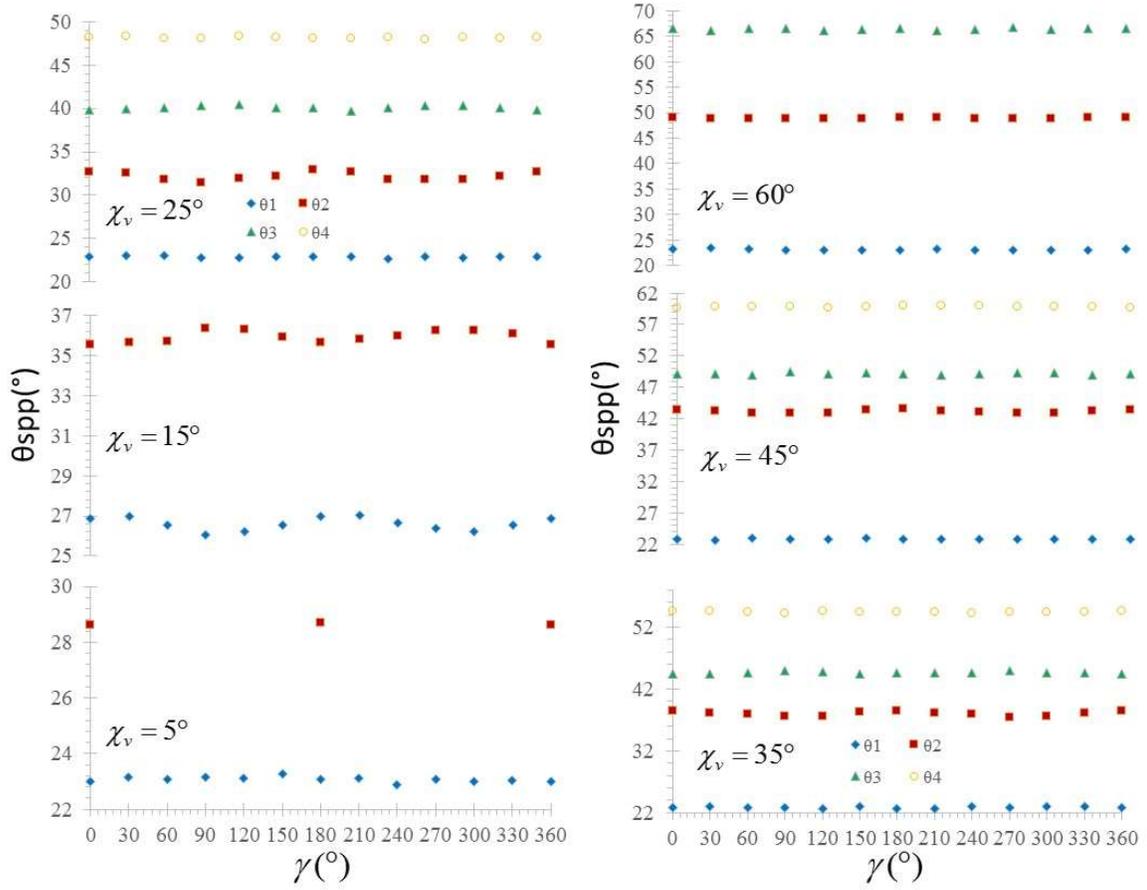

**Fig.2.** Multiple SPP peaks versus $\gamma$ for a titanium oxide ECSTF with a $d_{met}$ =10 nm central bulk aluminum layer ($\varepsilon_{met} = -56 + 21i$) for a p-linear polarized plane wave, when $\Omega$ =150 nm for $\chi_v \in \{5°, 15°, 25°, 35°, 45°, 60°\}$.

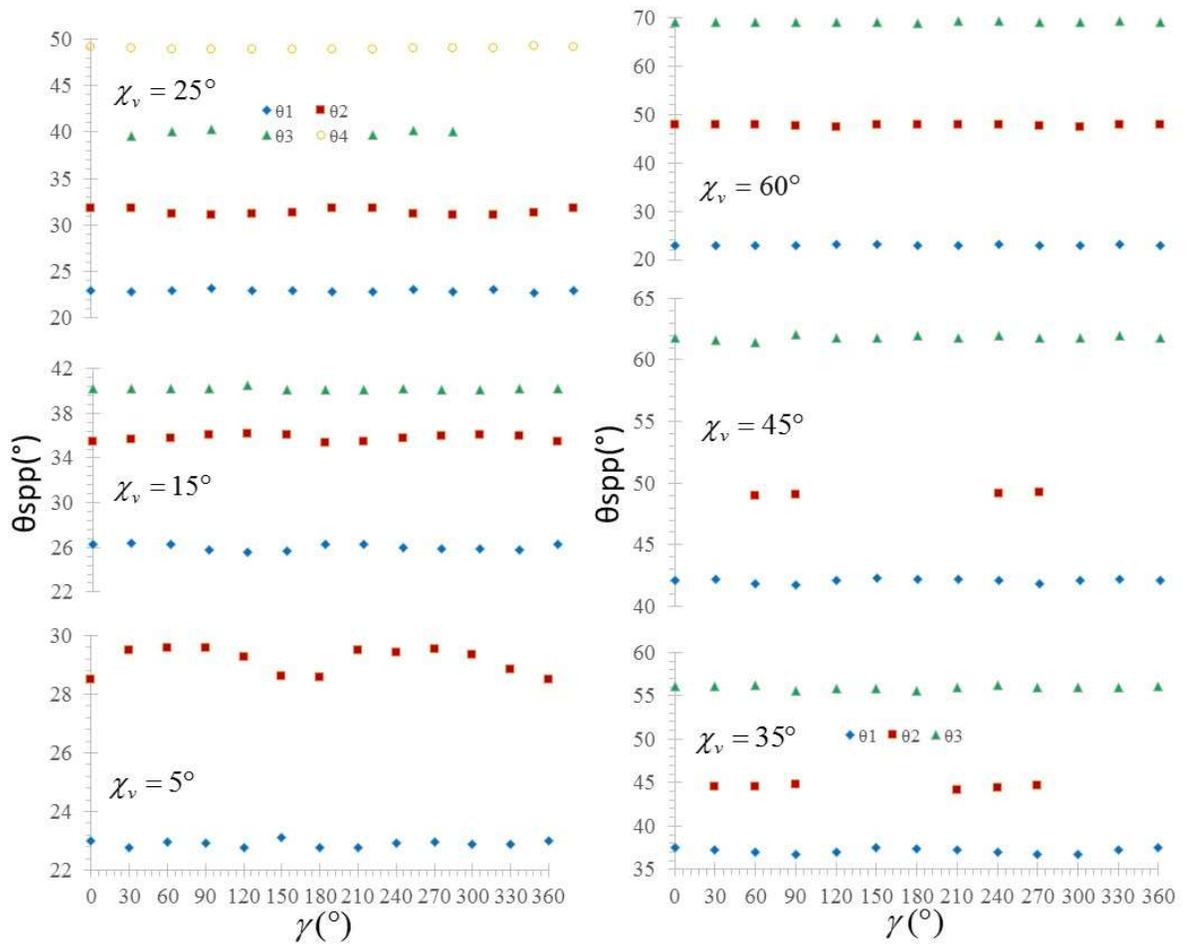

**Fig.3.** Same as Fig. 2, except that the $d_{met}$ =20 nm.

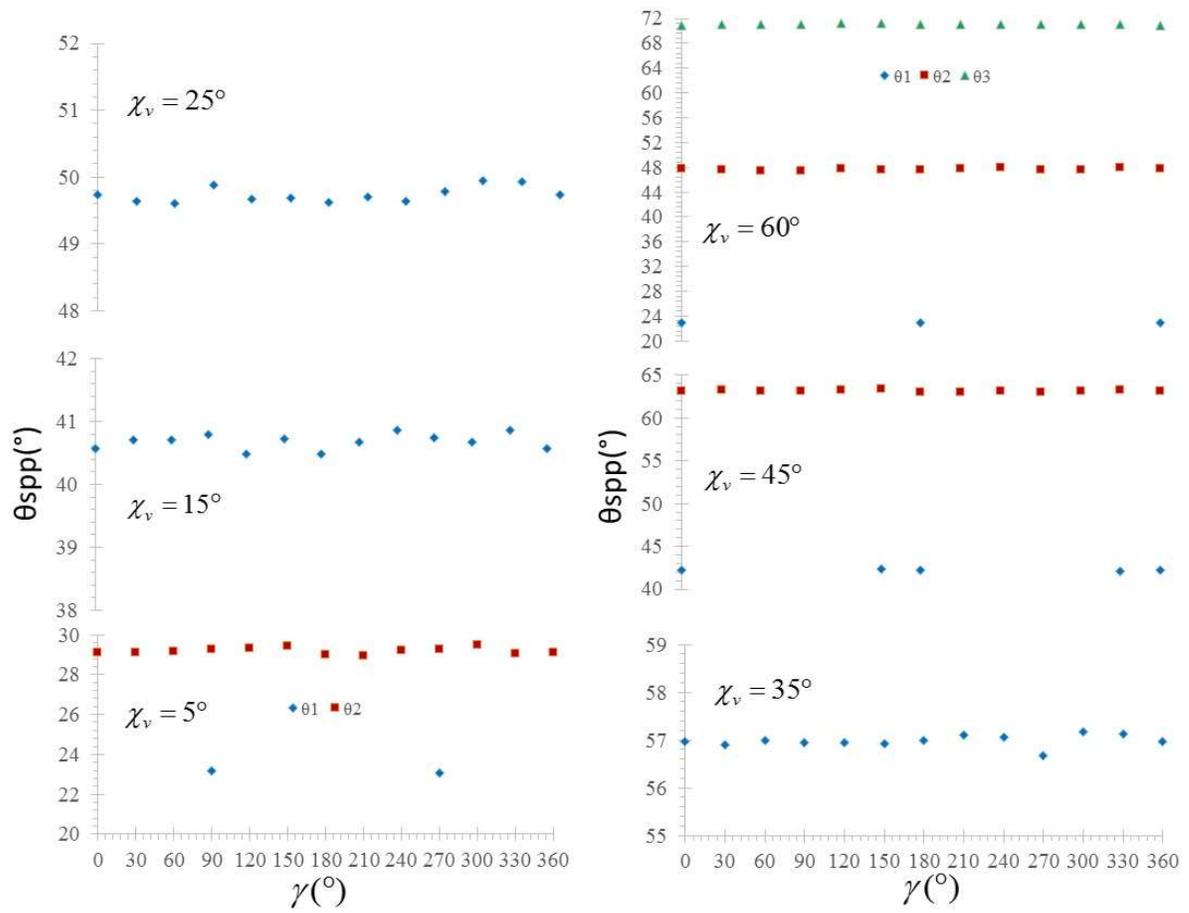

**Fig.4.** Same as Fig. 2, except that the $d_{met}$ =30 nm.

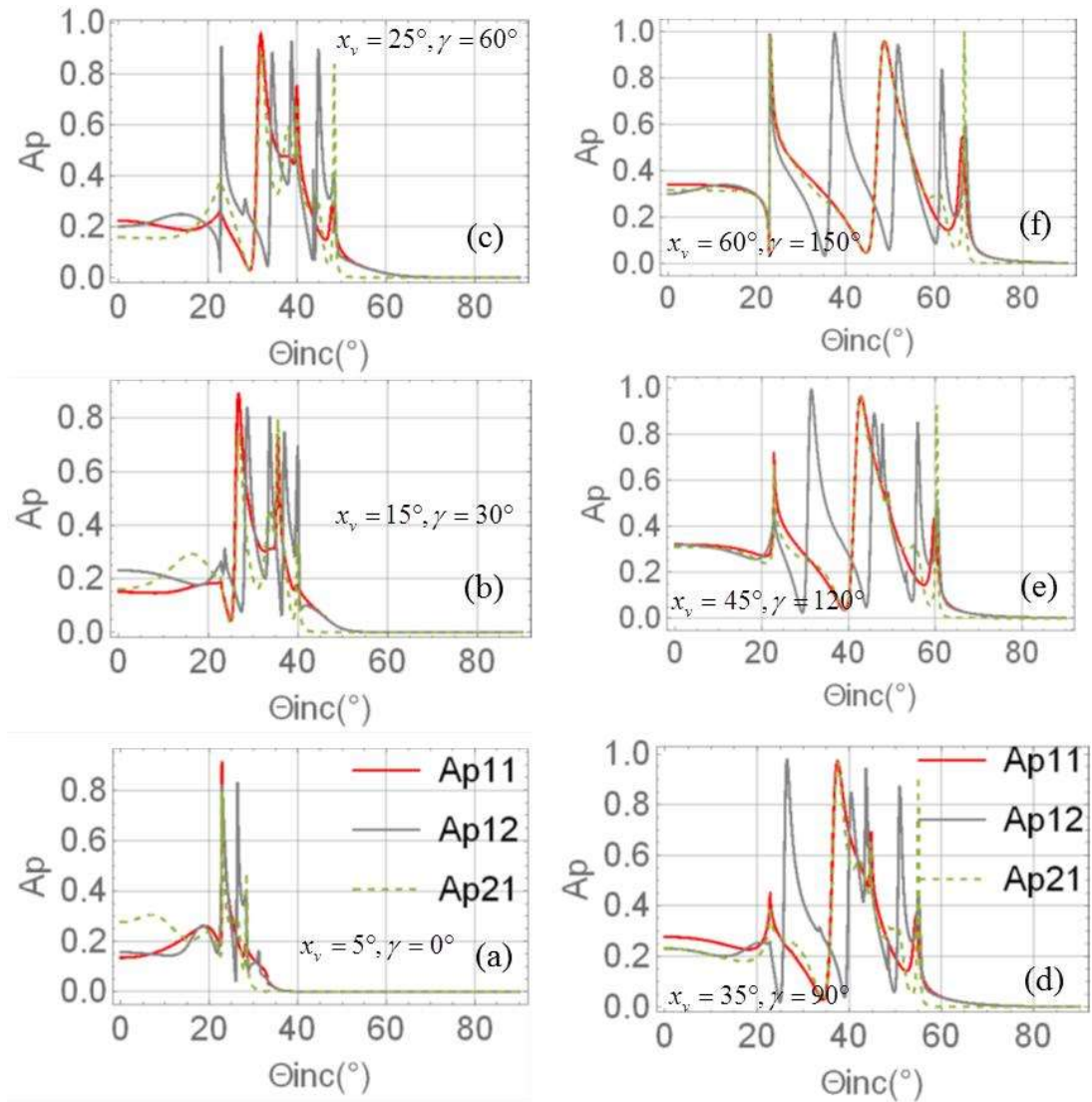

**Fig.5.** Optical absorption $Ap_{NbNt}$ as a function of incident angle of light for p-linear polarized plane wave at different vapor deposition and offset angles so that $N_b$ and $N_t$ are respectively the period of bottom and top of the ECSTF, when $\Omega$=150 nm, $d_{met}$= 10 nm.

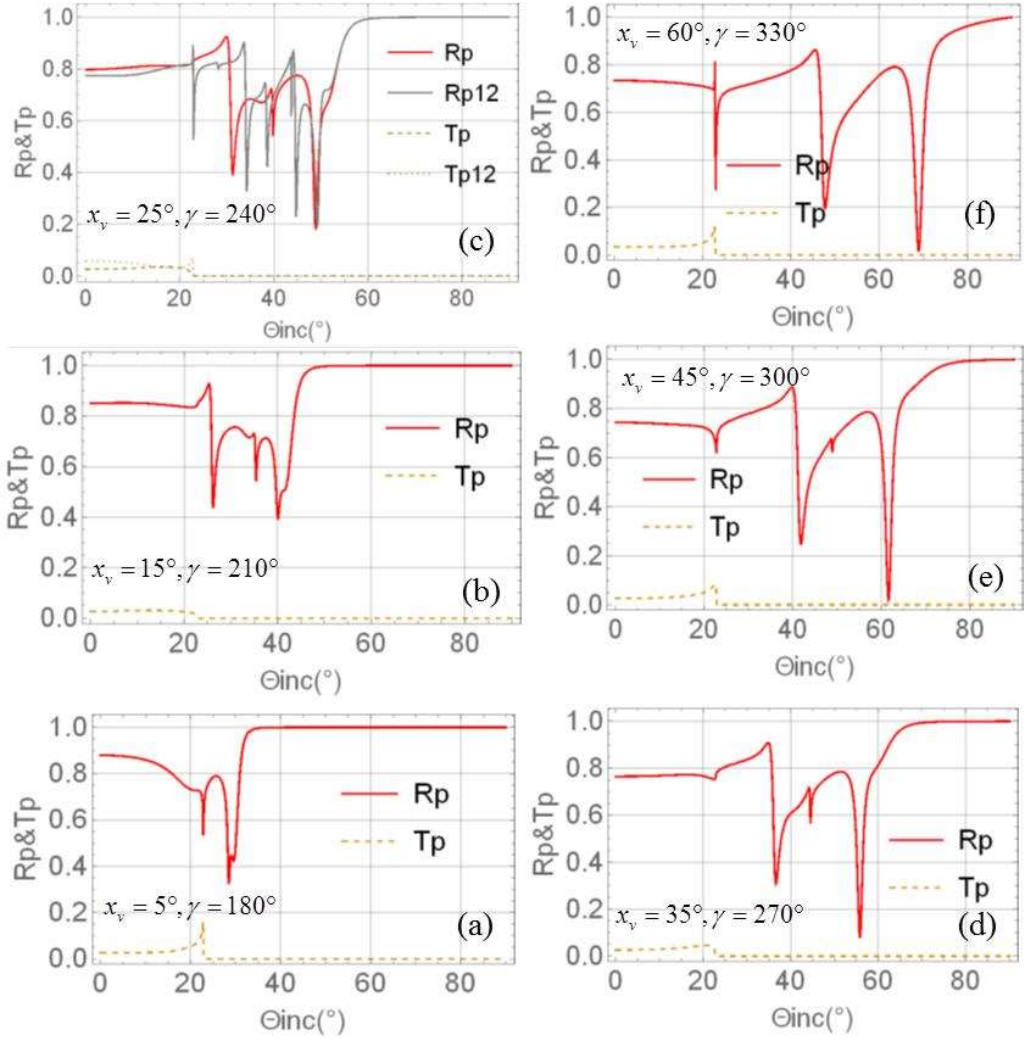

**Fig.6.** Reflection and transmission as a function of incident angle of light for p-linear polarized plane wave at different vapor deposition and offset angles, when $\Omega$ =150 nm, $d_{met}$ = 20 nm and $N_d = N_t =1$ except that it is mentioned.

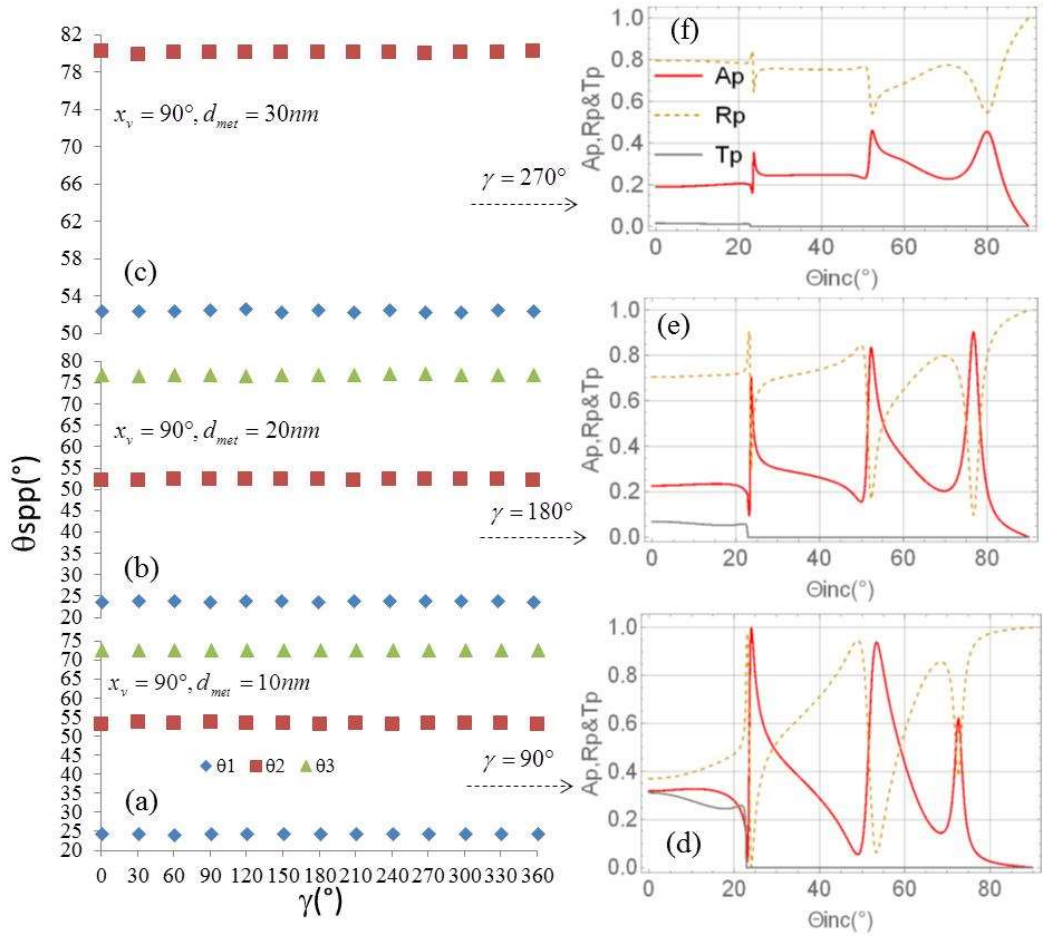

**Fig.7.** Multiple SPP peaks versus γ extracted from the optical absorption(Ap) at $\chi_v = 90°$ for $d_{met}$ =10,20,30 nm (a,b,c) and absorption, reflection and transmission of them drawn as a function of incident angle of light for p-linear polarized plane wave at γ = 90,180,270° (d,e,f).

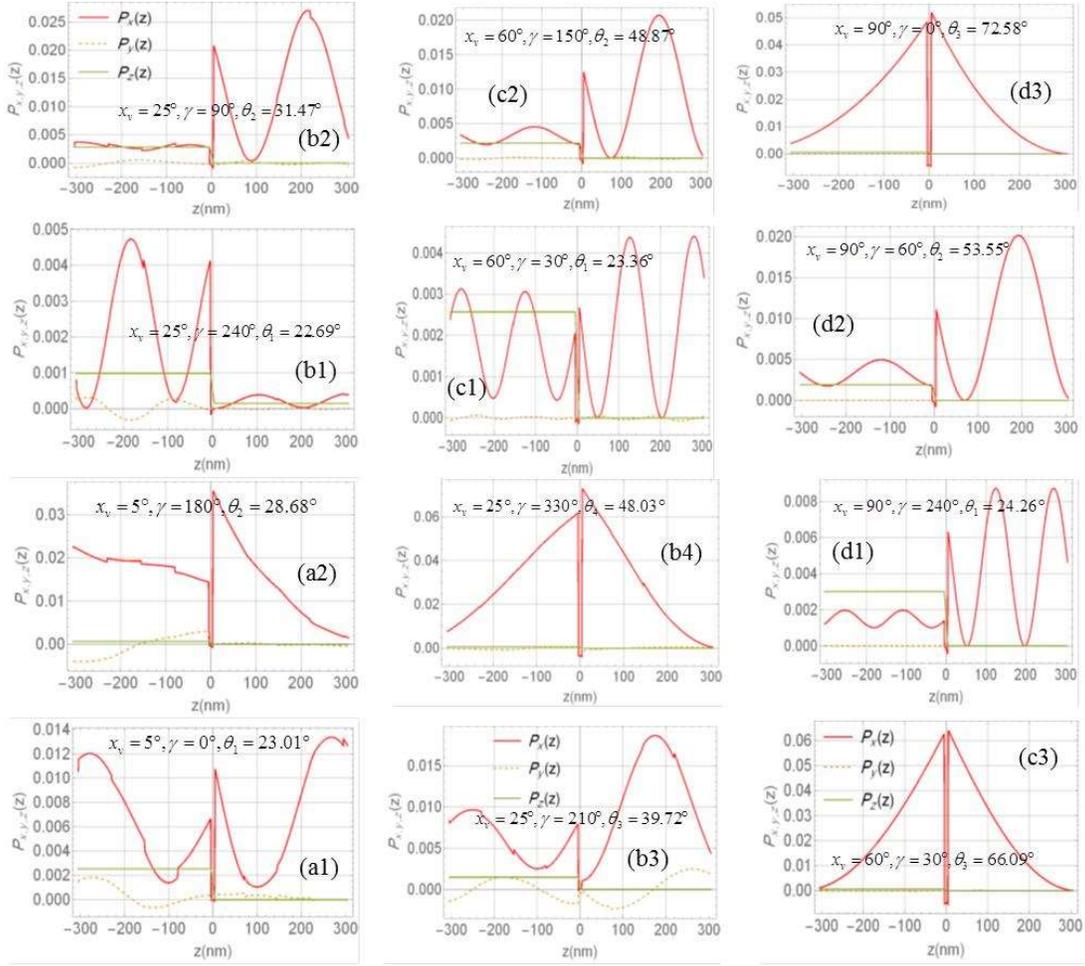

**Fig.8.** Cartesian components of the time-averaged Poynting vector $P_{x,y,z}(z)$ along the z axis at different vapor deposition and offset angles for p-linear polarized plane wave when a SPP wave has been excited for $d_{met}$ =10 nm. The *x*-, *y*- and *z*-directed components of P(*z*) are represented by solid red, dashed orange and solid green lines, respectively.

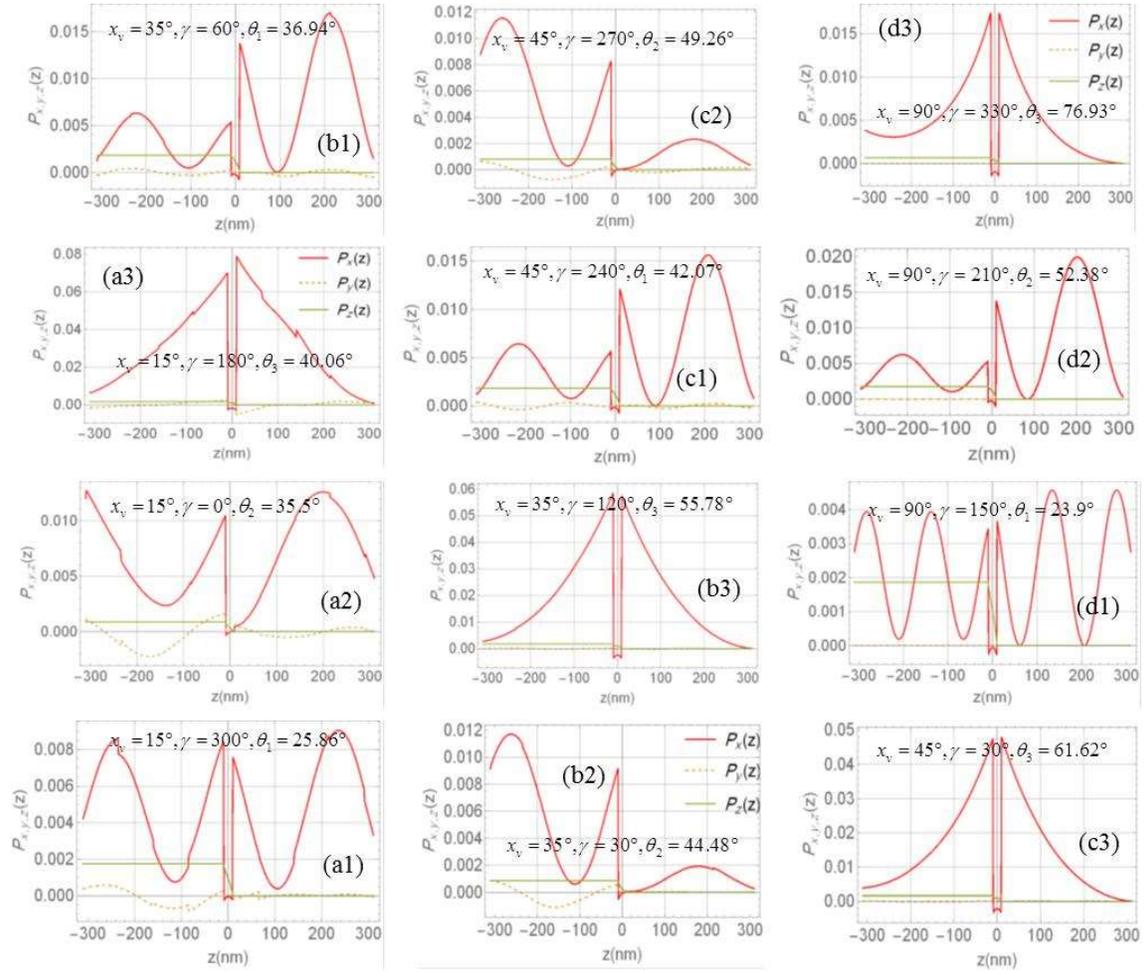

**Fig.9.** Same as Fig. 8, except that the $d_{met}$ =20 nm.

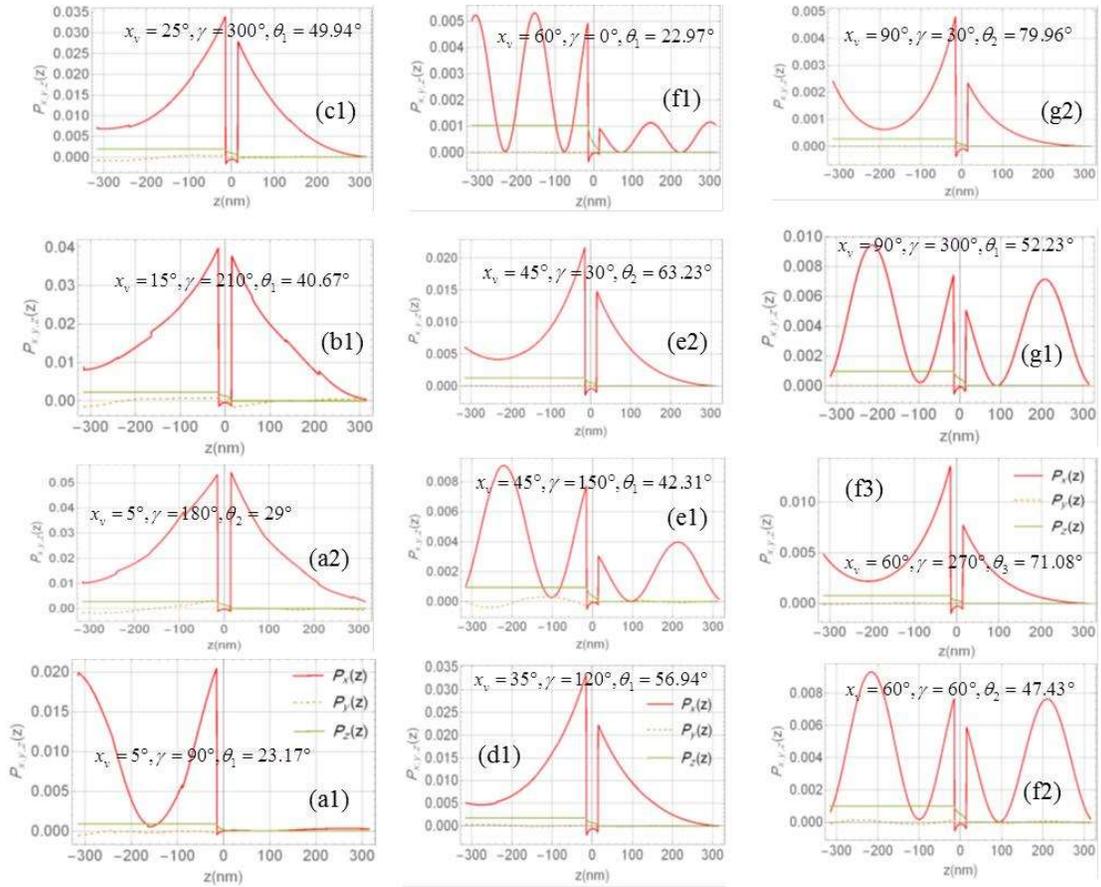

**Fig.10.** Same as Fig. 8, except that the $d_{met}$ =30 nm.